\documentclass{article}
\usepackage{ltwol2e,epsfig}
\usepackage{amssymb}

\def\EUP{{\em Eur.~Phys.~J.}~C}

\def\PRD{{\em Phys.~Rev.}~D}

\def\ab{\bar{\alpha}}
\def\be{\begin{equation}}
\def\ee{\end{equation}}
\def\bea{\begin{eqnarray}}
\def\eea{\end{eqnarray}}
\begin{document}
 
\pagestyle{empty}
 
\twocolumn[\begin{flushright}  
\large PITHA 98/36\\
\large hep-ph/9811210
\end{flushright}
\vskip3cm

\begin{center}
 
{\huge \bf QCD Tests in Hadronic Final States}
 
\vskip2.2cm
 
{\Large M. Wobisch}\\ \vskip5mm
\large III. Physikalisches Institut, RWTH Aachen \\
D-52056 Aachen, Germany
\end{center}
\vskip3cm

\begin{center}
\large \bf Abstract
\end{center}
\vskip3mm
\large 
\hskip11mm
\vbox{\hsize=15.6cm Various QCD studies
based on jet observables in hadronic final states in deep-inelastic 
scattering and hadron-hadron collisions are presented.
Measured quantities are event shape variables, jet rates and 
jet cross sections.
QCD analyses have been performed to determine the value of the 
strong coupling constant $\alpha_s(M_Z)$, the gluon density in the proton
and to test power corrections.}

\vskip6cm

\hskip11mm
\vbox{\hsize=15.6cm Talk given on behalf of the H1 and the ZEUS Collaboration
at the 29th International Conference
on High-Energy Physics (ICHEP '98), Vancouver, Canada, July 23--29 1998.}

\newpage
 \,
\newpage
] 

\pagestyle{plain}
\pagenumbering{arabic}

\title{QCD TESTS IN HADRONIC FINAL STATES}

\author{M. WOBISCH}

\address{III. Physikalisches Institut, RWTH Aachen,
D-52056 Aachen, Germany,\\
E-mail: Markus.Wobisch@desy.de}


\twocolumn[\maketitle\abstracts{Various QCD studies
based on jet observables in hadronic final states in deep-inelastic 
scattering and in hadron-hadron collisions are presented.
Measured quantities are event shape variables, jet rates and 
jet cross sections.
QCD analyses have been performed to determine the value of the 
strong coupling constant $\alpha_s(M_Z)$, the gluon density in the proton
and to test power corrections.}]

\section{Introduction}
Hadronic final states in high energy particle collisions 
offer a rich field for testing quantum chromodynamics (QCD).
Especially suited for such studies are infrared and collinear safe 
``jet observables" for which QCD predictions can be 
calculated perturbatively.
In this contribution measurements are presented of
event shape variables~\cite{H1evtshp,Zeusevtshp}, 
jet rates~\cite{H1y2,H1dijrat} and jet cross sections~\cite{H1dijet}
in deep-inelastic scattering (DIS) by the H1 and the ZEUS collaboration
as well as jet cross sections
in proton-antiproton collisions~\cite{CDFincl} by the CDF collaboration.
These observables are directly (i.e.\ already at leading order)
sensitive to the value of the strong coupling constant $\alpha_s$.
Furthermore the jet cross sections are directly sensitive to the 
gluon density in the proton.
For all these observables, the perturbative cross sections have been
calculated to next-to-leading order (NLO) in $\alpha_s$.
Based on these data, QCD analyses have been performed to extract the
value of the strong coupling constant $\alpha_s(M_Z)$ and to determine the 
gluon density in the proton.
The event shape data have been analyzed for tests of ``power corrections".

\section{Event Shape Variables in DIS}
\begin{figure}[t]
\center
\epsfig{file=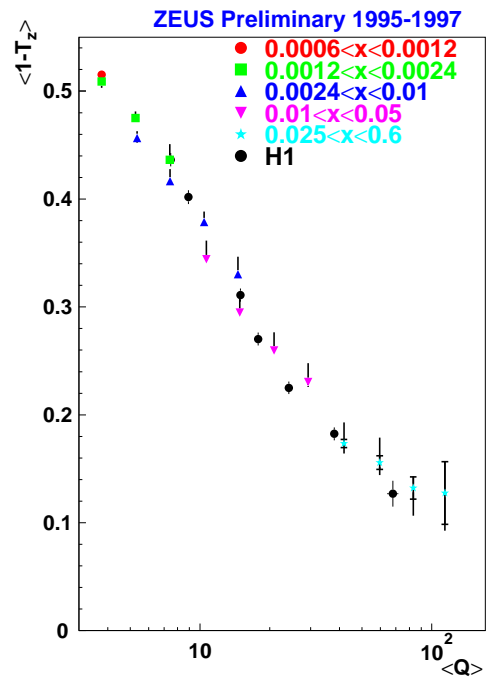,height=9cm,clip=}
\caption{The $Q$ dependence of the mean values of $1-T_Z$ measured in 
deep-inelastic scattering in the current hemisphere of the Breit frame.
The ZEUS data are measured in different bins of $x_{\rm Bj}$ while the H1 
results are integrated over all $x_{\rm Bj}$.}
\label{fig:zeus_shapes}
\end{figure}

\begin{figure}[t]
\center
\epsfig{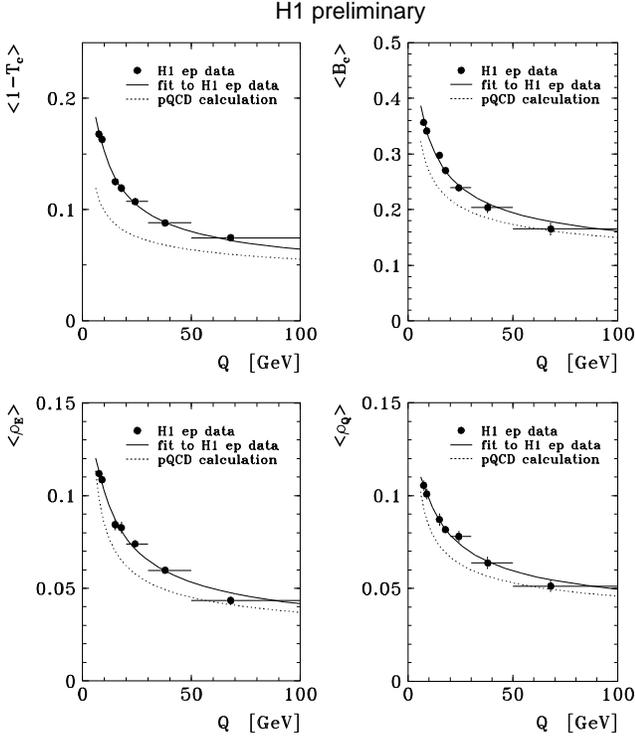}
\caption{Event shape variables defined in the current hemisphere
of the Breit frame in deep-inelastic scattering.
Displayed is the $Q$ dependence of the mean values of the different 
event shapes.
The data are compared to the results of a QCD fit involving 
NLO predictions combined with power corrections.
Also shown is the NLO prediction.}
\label{fig:evtshapes}
\end{figure}

Event shape variables in deep-inelastic scattering events have been 
investigated over a large range of momentum transfers 
$4 < Q < 100\,\mbox{GeV}$, based on data taken at HERA by the 
H1 experiment~\cite{H1evtshp} in 1994--1996 and by the 
ZEUS experiment~\cite{Zeusevtshp} in 1995--1997.
These event shapes are defined by linear sums of momenta of all 
hadronic final state particles in the current hemisphere of the Breit frame
and are in different ways sensitive to the collimation of the energy flow 
along the event shape axis:
\begin{eqnarray*}
\mbox{Thrust:} \hskip2mm & T_z & = 
\frac{\sum p_\|}{\sum | \vec{p} |} \hskip4mm (\tau_z = 1 - T_z) \, , \\
  & T_c & = \max_{\vec{n_T}} 
 \frac{\sum |\vec{p} \cdot \vec{n_T} | }{\sum | \vec{p} |}
  \hskip3mm (\tau_c = 1 - T_c) \, , \\
\mbox{Jet Broadness:}\hskip2mm & B_c & = 
\frac{\sum p_\perp}{\sum | \vec{p} |} \, , \\
\mbox{Jet Mass:}\hskip2mm & \rho_E  & = 
\frac{(\sum p)^2}{2\,(\sum E)^2} \, , \; \; \\
    & \rho_Q &  = \frac{(\sum p)^2}{Q^2} \, .
%     \rho_Q   = \frac{(\sum p)^2}{Q^2} \, .
\end{eqnarray*}
While in the analysis by the H1 collaboration only the dependence 
on the four-momentum transfer $Q$ is measured (integrating over the
Bjorken scaling variable $x_{\rm Bj}$), 
the ZEUS collaboration has also studied the dependence on $x_{\rm Bj}$.
The mean values of $\tau_z$ are shown in Fig.~\ref{fig:zeus_shapes}
as a function of $Q$. 
The $Q$ dependence of the mean values of the other event shapes
is shown in Fig.~\ref{fig:evtshapes}.
It can be seen that at larger values of $Q$ and $x_{\rm Bj}$
the final state in the current hemisphere of the 
Breit frame is more collimated (relative to its total energy).

\noindent
The QCD prediction for the mean value of any event shape $\left< F \right>$ 
can be written as a sum of perturbative and non-perturbative contributions:

\bea
& \left< F \right> & = \; \left< F \right>^{\mbox{\scriptsize pert}} 
\;+\;\left< F \right>^{\mbox{\scriptsize pow}}   , \\
\mbox{with} \nonumber \\
& \left< F \right>^{\mbox{\scriptsize pert}}  & = \;
c_1 \; \alpha_s(Q) \; + \; c_2 \; \alpha^2_s(Q) \,, \nonumber \\
& \left< F \right>^{\mbox{\scriptsize pow}} & = \; a_F \, \frac{16}{3\pi}
 \, \frac{\mu_I}{Q} \,  ( \ab_0(\mu_I) - \alpha_s(Q)  
- b_0 \alpha^2_s(Q) ) \, . \nonumber 
\eea
The perturbative part can be calculated to next-to-leading order.
Apart from the dependence on the parton density functions (included in the
coefficients $c_i$) the only free parameter is $\alpha_s$.
The non-perturbative contributions are modeled by
$a_F/Q$ power corrections, where the $a_F$ have been calculated~\cite{powcor}.
This contribution also depends on the value of 
$\bar{\alpha}_0(\mu_I)$, the effective coupling below the 
infrared matching scale $\mu_I$.

\begin{figure}[t]
\center
\epsfig{file=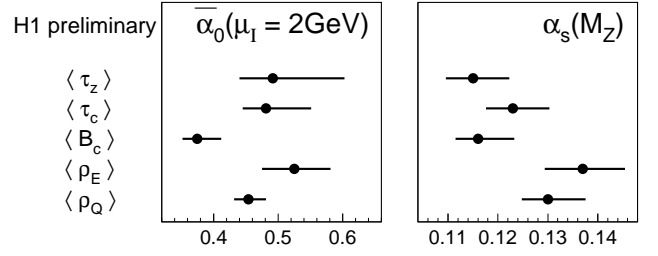,width=8.65cm}
\caption{The results for $\bar{\alpha}_0$ and $\alpha_s$ as determined
in a fit of the theoretical predictions based on NLO pQCD and 
power corrections to the $Q$ dependence of the event shape mean values.}
\label{fig:a0_as_evtshp}
\vskip0.5mm
\end{figure}

Fits have been performed to the mean values of the event shapes with the 
free parameters $\alpha_s(M_Z)$ and $\bar{\alpha}_0(\mu_I)$ 
according to eq.~(1). 
The fit results in Fig.~\ref{fig:evtshapes} show smaller non-perturbative
contributions towards higher $Q$.

\noindent
The resulting values of the strong coupling constant $\alpha_s(M_Z)$ 
and $\bar{\alpha}_0$ (displayed in Fig.~\ref{fig:a0_as_evtshp})
are of the same size and demonstrate approximately universal behavior
for all event shapes.
Remaining discrepancies may be traced back to large theoretical
uncertainties: large NLO corrections and a large renormalization scale 
dependence indicate possible large contributions from higher, 
uncalculated orders.

\section{Multijet Rates in DIS}
To determine $\alpha_s$, the production rate of multijet events in DIS 
has been studied based on data taken in 1994 and 1995 by the H1 collaboration.
Two analyses are presented where jets are defined by the modified JADE 
jet algorithm~\cite{mjade} (mJADE), applied in the HERA laboratory frame.
In the mJADE jet definition the final state particles (and a pseudoparticle
to account for the non-visible proton remnant) are clustered in the order
of increasing invariant di-particle masses.
For this jet definition the jet resolution parameter $y_{ij}$ is given by 
$\, y_{ij} \; = \; M^2_{ij} \; / \; W^2 $,
where $W$ is the total invariant mass of the hadronic final state
and $M_{ij}$ is the smallest invariant mass between pairs of
\mbox{(pseudo-)} particles.

\begin{figure}
\center
\epsfig{file=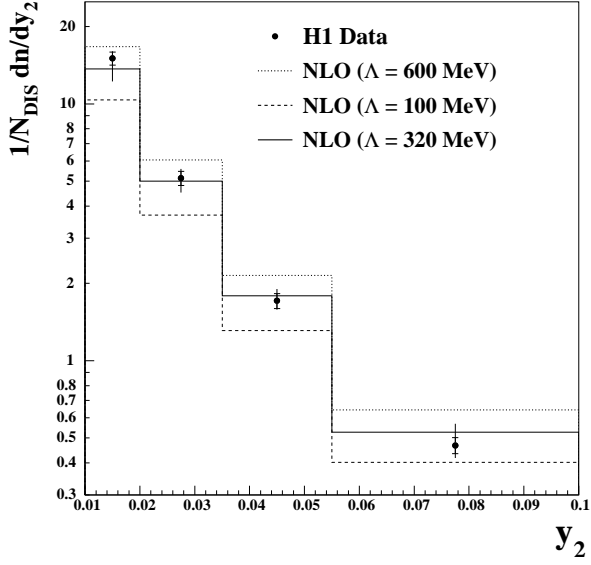,width=8.2cm}
\vskip-3mm
\caption{The $y_2$ spectrum measured in deep-inelastic scattering with
the mJADE algorithm: The data, corrected to ``parton-level", are 
compared to the NLO prediction for different values of 
$\Lambda^{(4)}$ using MRSH parton distributions. 
The best fit result is obtained for the value of 
$\Lambda^{(4)} = 320\,\mbox{MeV}$ 
corresponding to a value of $\alpha_s(M_Z) = 0.118$.}
\label{fig:as_y2}
\end{figure}

\begin{figure}
\center
\epsfig{file=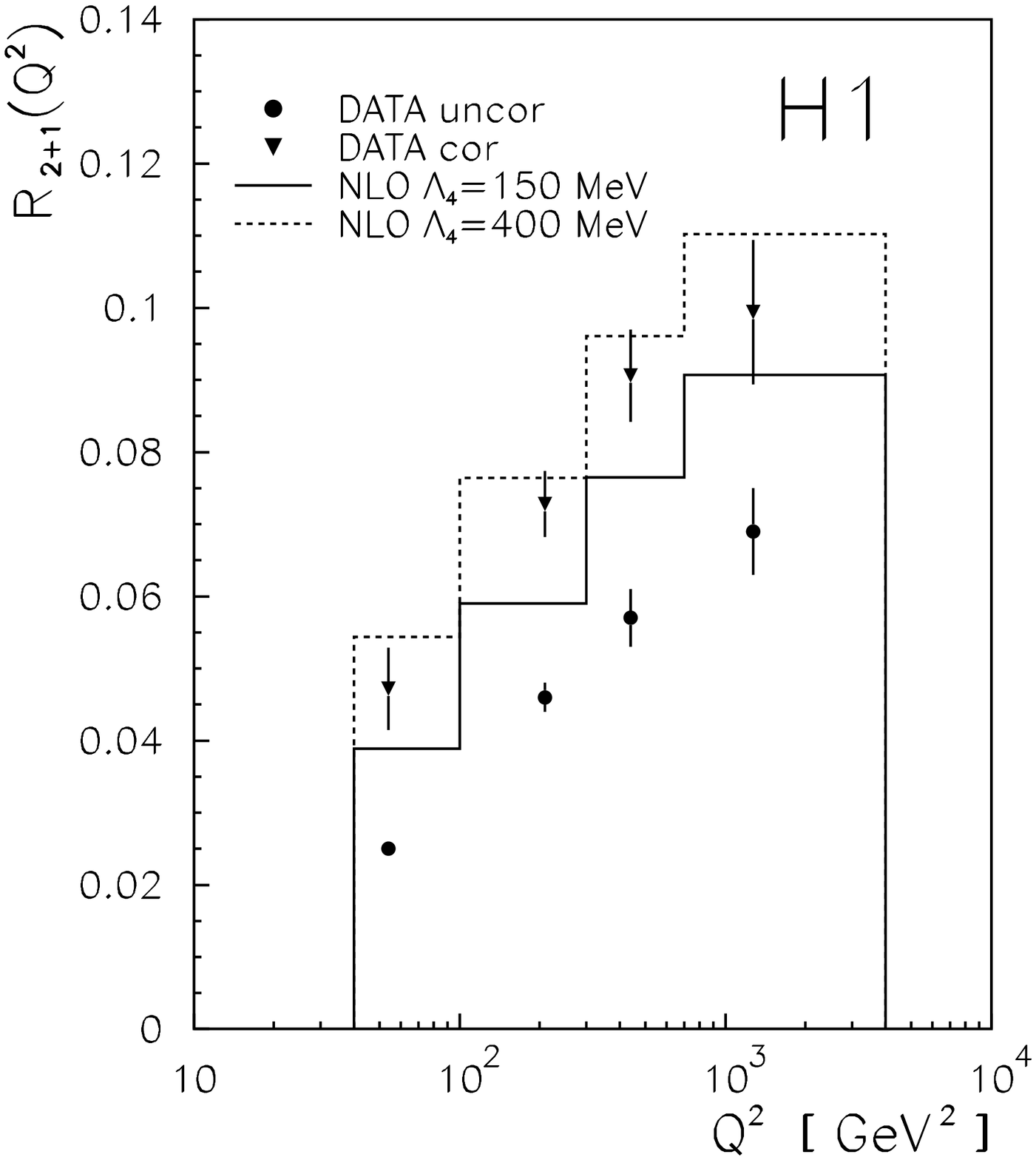,width=7.9cm,clip=}
\epsfig{file=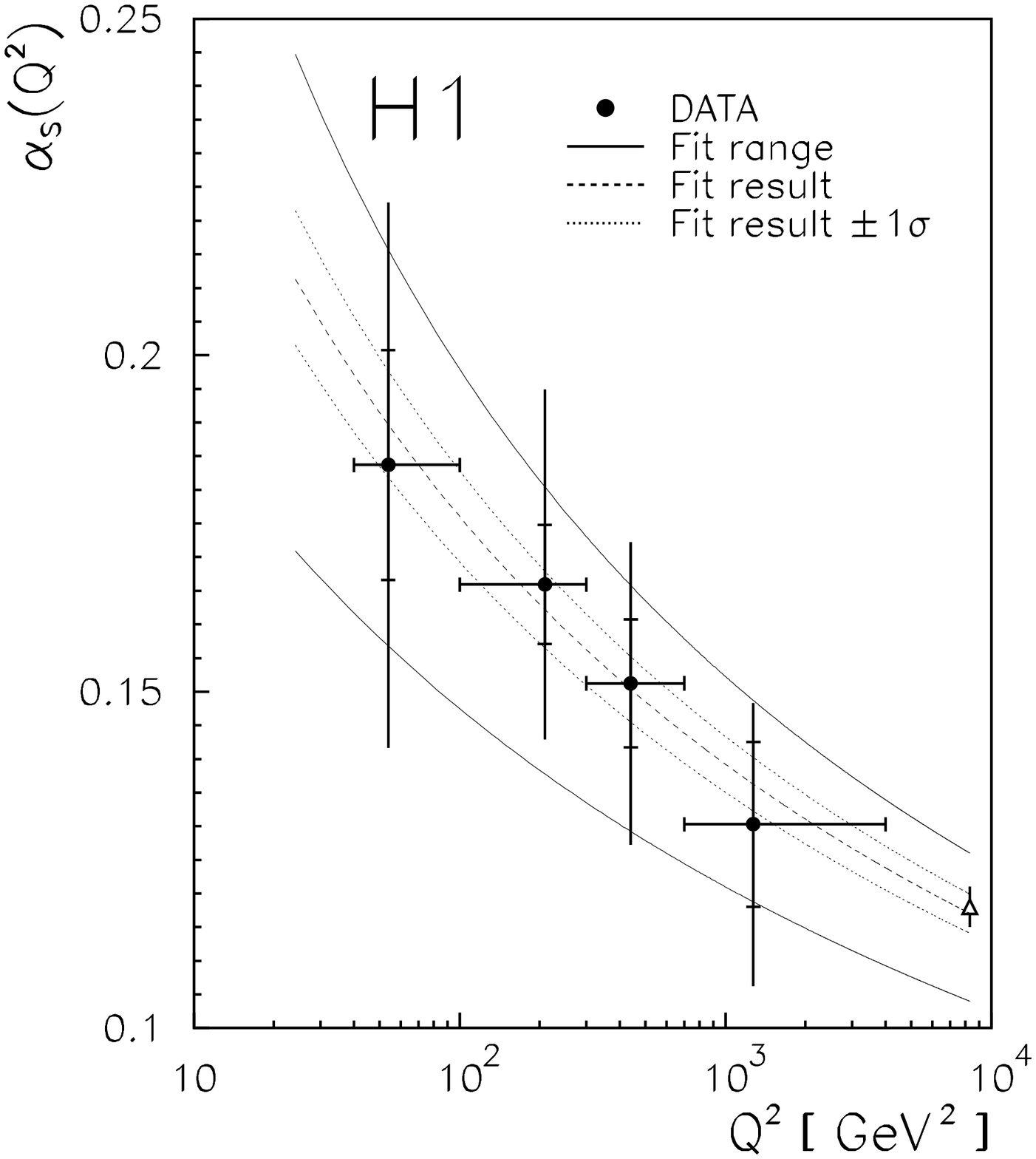,width=7.9cm,clip=}
\caption{The rate of (2+1)-jet events in in deep-inelastic scattering
as defined by the mJADE algorithm. The upper plot shows the data, 
corrected to ``parton-level" together with the NLO predictions for 
different values of $\Lambda^{(4)}$ for MRSH parton distributions.
The lower plot shows the values of the strong coupling constant 
$\alpha_s$ extracted in each $Q^2$ bin.
The inner error bars show the statistical error; the total error
is given by the full error bars.
The result of the fit of the renormalization group equation to the single 
$\alpha_s(Q^2)$ values (considering only the statistical errors)
is shown as the dashed line, the errors as dotted lines.
Taking also into account the full systematic uncertainties results
in a fit range which is displayed by the full lines.
The central fit result corresponds to a value of $\alpha_s(M_Z) = 0.117$.}
\label{fig:as_dijrat}
\end{figure}

Two observables have been measured: 
the $y_2$ spectrum~\cite{H1y2} and the (2+1)-jet rate~\cite{H1dijrat}, 
the latter as a function of $Q^2$.
The variable $y_2$ is defined as the value of the jet resolution 
parameter $y_{ij}$ at which exactly (2+1) jets are resolved 
(the ``+1" denotes the proton remnant).
The $y_2$ spectrum has been measured for momentum transfers
$200 < Q^2 < 10000\,\mbox{GeV}^2$ (Fig.~\ref{fig:as_y2}).
The (2+1)-jet rate $R_{2+1}$ is here defined as the fraction of DIS events 
that have (2+1) jets at a fixed value of the jet resolution parameter 
$y_{\mbox{\scriptsize cut}} = 0.02$ (i.e.~the rate of events which have
(2+1) jets after the clustering has been stopped when all 
$M^2_{ij} / W^2 > 0.02$).
The results have been obtained in four $Q^2$ bins 
at $40< Q^2< 4000\,\mbox{GeV}^2$ (Fig.~\ref{fig:as_dijrat} top).

Both measurements have been used to extract $\alpha_s(M_Z)$.
For the $y_2$ spectrum, a fit has been performed to determine $\alpha_s$ at a
scale of $\mu^2_r = \langle Q^2 \rangle = 620\,\mbox{GeV}^2$.
Using the two-loop formula, this value is evolved to $\mu^2_r = M^2_Z$.
For the jet rate $R_{2+1}(Q^2)$, values of $\alpha_s$ have been 
determined in each $Q^2$-bin at a scale $\mu^2_r = Q^2$ 
(Fig.~\ref{fig:as_dijrat} bottom).
A combined fit of the two-loop formula to these four values 
leads to the result for $\alpha_s(M_Z)$. 
The results from both analyses are
\begin{eqnarray*}
y_2: \;  &\alpha_s(M_Z) =  & 0.118 \pm 0.002  \mbox{\small (stat)}
\,^{+0.007}_{-0.008} \mbox{\small (syst)}   \\
 &&\hskip9mm \,^{+0.007}_{-0.006} \mbox{\small (theor)} \, , \\
R_{2+1}:\; &\alpha_s(M_Z) =&  0.117 \pm 0.003 \mbox{\small (stat)}
 \,^{+0.009}_{-0.013} \mbox{\small (syst\&theor)}  \\
&& \hskip8mm \; +0.006 \mbox{\small (rec.scheme)} \, .
\end{eqnarray*}

\noindent
The extracted $\alpha_s(M_Z)$ values are in good agreement with 
each other and with the current world average value. 
For both results the largest uncertainties come from a variation
of the parton densities used, the renormalization scale dependence 
of the NLO prediction and from a large model dependence of the size 
of the non-perturbative hadronization corrections.

\section{\boldmath Inclusive Jet Cross Section in $\bar{p}p$ Collisions}

\begin{figure}
\center
\epsfig{file=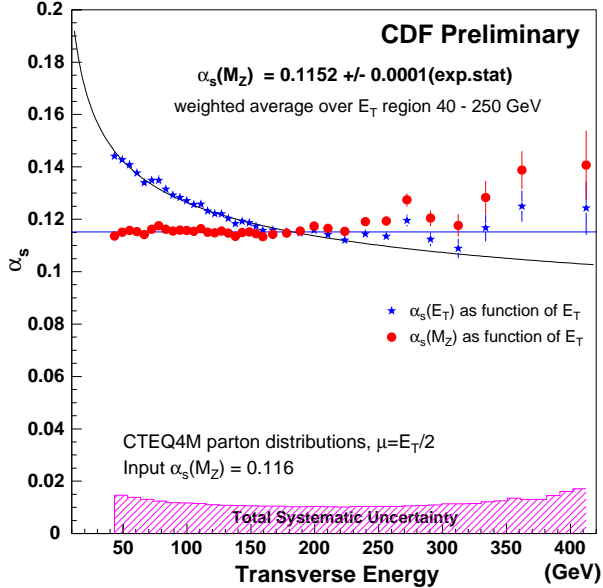,width=8.2cm}
\caption{The values of the strong coupling constant $\alpha_s(E_T/2)$
and $\alpha_s(M_Z)$ as determined from the inclusive jet cross section 
in $\bar{p}p$-collisions. 
The single values have been evolved to $\mu=M_Z$ using the two-loop formula.
The error weighted average over the $E_T$ range 40 -- 250\,GeV gives
the final result  of $\alpha_s(M_Z) = 0.115$.}
\label{fig:incljet_as}
\vskip-1mm
\end{figure}

Inclusive jet production in $\bar{p}p$ collisions has been studied 
by the CDF collaboration based on data taken at the Tevatron during Run 1B.
The inclusive jet cross section has been measured~\cite{elvira} 
over a wide range of transverse jet energies 
$40 < E_T < 450\,\mbox{GeV}$ using a cone jet algorithm with a 
cone size of $R=0.7$ in the pseudo-rapidity range between 0.1 and 0.7.

\noindent
Each of the 36 individual bins in $E_T$ allows to extract a value of 
$\alpha_s$ at a scale chosen to be $\mu = E_T / 2$~\cite{CDFincl}.
The single $\alpha_s(E_T/2)$ values have been evolved to $\alpha_s(M_Z)$
using the two-loop equation (Fig.~\ref{fig:incljet_as}).
By forming an error weighted average over the $E_T$ range 
$40 - 250\,\mbox{GeV}$, an average value of $\alpha_s(M_Z)$ has been obtained
(the exclusion of the points $E_T > 250\,\mbox{GeV}$ avoids possible
bias by the population of events at high $E_T$).
For the CTEQ4M parton distributions, the result is

$\alpha_s(M_Z) = 0.1152 \pm0.0001 \mbox{(stat.)} \,^{+0.0083}_{-0.0093}
\mbox{(exp.syst.)}$ 

\noindent
with a theoretical uncertainty of
$\pm 0.005$ from the renormalization scale dependence~\cite{mudep}.

\begin{figure}
\center
\vskip1mm
\epsfig{file=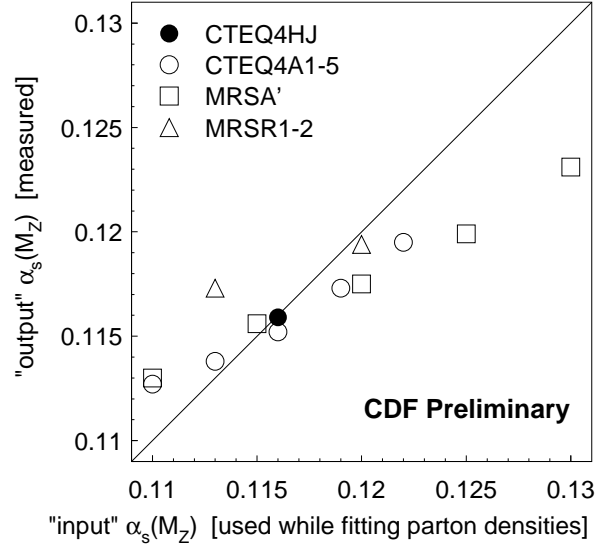,width=8cm}
\caption{The values of the strong coupling constant $\alpha_s(M_Z)$
as derived from a QCD fit to the inclusive jet cross section in 
$\bar{p}p$-collisions.
Several fits have been performed for different parameterizations
of parton density functions.
Shown is the dependence of the central fit result of $\alpha_s(M_Z)$
on the value of $\alpha_s(M_Z)$ used in the fit of the
parton distributions.}
\label{fig:as_corr}
\end{figure}

The result also depends on the parton densities used 
in the NLO calculation.
This dependence is demonstrated in Fig.~\ref{fig:as_corr}
where the resulting $\alpha_s(M_Z)$ value is shown as a function
of the $\alpha_s(M_Z)$ used in the global fit of the parton densities.
A clear correlation is seen which reflects the dependence of the 
inclusive jet cross section on the gluon density $g(x)$ in the proton.
The size of the gluon density in different global fits is 
anti-correlated to the assumed value of $\alpha_s$, since most 
cross sections only depend on the product $\alpha_s \cdot g(x)$.

Due to the dependence of the extracted $\alpha_s(M_Z)$ on the parton
densities (and their $\alpha_s$ value) this analysis can not be seen
as an independent measurement.
However, the fact that for standard parton distributions (as CTEQ4M)
the extracted $\alpha_s$ is close to the $\alpha_s$ used in the parton
densities is an important demonstration of consistency.

\section{Dijet Cross Sections in DIS}
To determine the gluon density in the proton, dijet production 
in deep-inelastic scattering has been studied by the 
H1 collaboration based on data taken in 1994--1997~\cite{H1dijet}.
The dijet cross section has been measured for different 
$k_\perp$-type jet algorithms in the Breit frame.
QCD models predict that hadronization corrections are smallest for
the longitudinally boost-invariant $k_\perp$ jet algorithm~\cite{invkt}
(about 7\% with negligible model dependence).

For this algorithm the inclusive dijet cross section has been measured
as a function of several observables: $Q^2$, $x_{\rm Bj}$, 
the average transverse jet energy of the two jets in the Breit frame 
$\overline{E}_T$, the invariant dijet mass $M_{jj}$ and the variable
$\xi = x_{\rm Bj} (1+ M^2_{jj}/Q^2)$ which is (at leading order) equal
to the momentum fraction of the proton carried by the parton that enters 
the hard process.
Dijet events have been selected where at least two jets have transverse 
energies of $E_T > 5\,\mbox{GeV}$ in the Breit frame. 
In addition, the sum of the transverse jet energies of the two 
highest $E_T$ jets in the event has been required 
to be $E_{T1}+E_{T2} > 17\,\mbox{GeV}$.

\begin{figure}[t]
\center
\epsfig{file=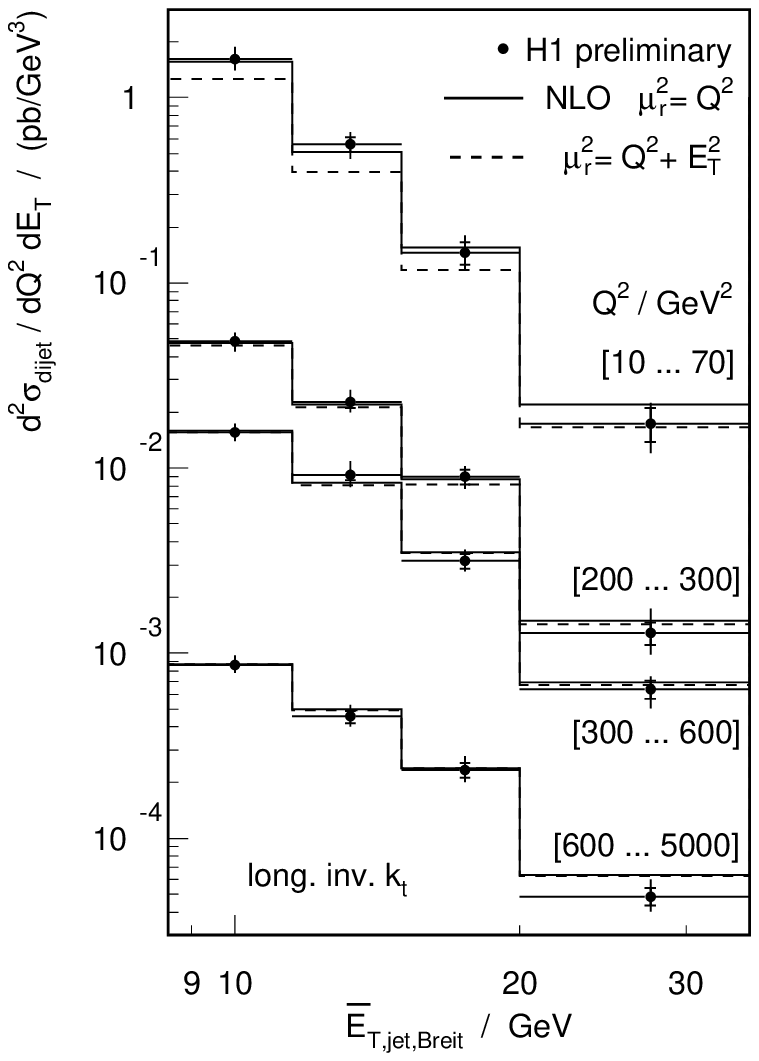,width=7.7cm}
\vskip-2mm
\caption{The double differential dijet cross sections
${\rm d}^2 \sigma / {\rm d}Q^2 {\rm d}\overline{E}_T$ 
in deep-inelastic scattering.
The data are compared to the NLO prediction for two different choices 
of the renormalization scale for CTEQ4M parton distributions.}
\label{fig:et}
\end{figure}

\begin{figure}[t]
\center
\epsfig{file=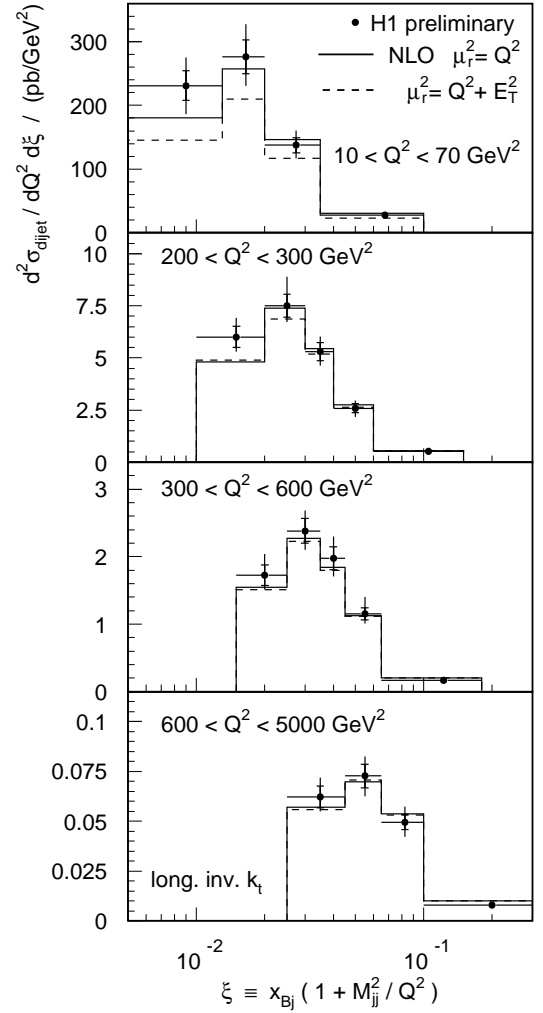,width=7.3cm}
\vskip-2mm
\caption{The double differential dijet cross sections
${\rm d}^2 \sigma / {\rm d}Q^2 {\rm d}\xi$ in deep-inelastic scattering.
The data are compared to the NLO prediction for two different choices 
of the renormalization scale for CTEQ4M parton distributions.}
\label{fig:xi}
\end{figure}

The dijet cross section is shown in different bins of $Q^2$ as a 
function of $\overline{E}_T$ (Fig.~\ref{fig:et}) and the 
variable $\xi$ (Fig.~\ref{fig:xi}) for $10 < Q^2 < 5000\,\mbox{GeV}^2$.
The data are compared to the NLO prediction for different
renormalization scales.
Reasonable agreement is observed, except at smaller values of $\xi$
where the NLO prediction is slightly too low.

\pagebreak

\noindent
The determination of the gluon density in a QCD fit to these 
dijet cross sections also requires knowledge of the quark densities 
and of $\alpha_s$.
In the present analysis $\alpha_s(M_Z)$ has been 
taken to be the world average~\cite{worldas}
within its uncertainty ($\alpha_s(M_Z) = 0.119 \pm 0.005$)
and subsequently evolved to the relevant scales using the two-loop equation.
To have a consistent treatment of the quark densities in the QCD fit,
H1 data on the inclusive neutral current DIS cross 
section~\cite{H1incl} at $200 < Q^2 < 650\,\mbox{GeV}^2$ have been
included in the fit.
These data give strong constraints on the quark densities.

The NLO corrections to the dijet cross sections at 
$Q^2 \lesssim 100\,\mbox{GeV}^2$ are sizeable (50 -- 100\,\%) and
the renormalization scale dependence is large (up to 20\,\%).
To avoid these theoretical uncertainties, only dijet data at 
$Q^2 > 200\,\mbox{GeV}^2$ have been considered in the fit.
In this range, approx.~55\,\% of the dijet cross section is gluon induced.
To maximize the sensitivity to the $x$-dependence of the gluon density,
the fit has been performed on the double differential cross sections 
${\rm d}^2\sigma / {\rm d}Q^2 {\rm d} \xi$ and 
${\rm d}^2\sigma / {\rm d}Q^2 {\rm d} x_{\rm Bj}$.
The gluon and quark densities have been fitted at a factorization scale
of $\mu^2_f= 200\,\mbox{GeV}^2$ which is of the size of the hard scales
for both the dijet and the inclusive cross section
(dijet cross section: $\mu^2_f \simeq \overline{E}^2_T$,
inclusive cross section: $\mu^2_f \simeq Q^2$).
The $x$-dependence of the gluon and quark densities has been parameterized
by the usual 3-, 4- or 5-parameter functions~\cite{H1dijet}.

\begin{figure}
\center
\epsfig{file=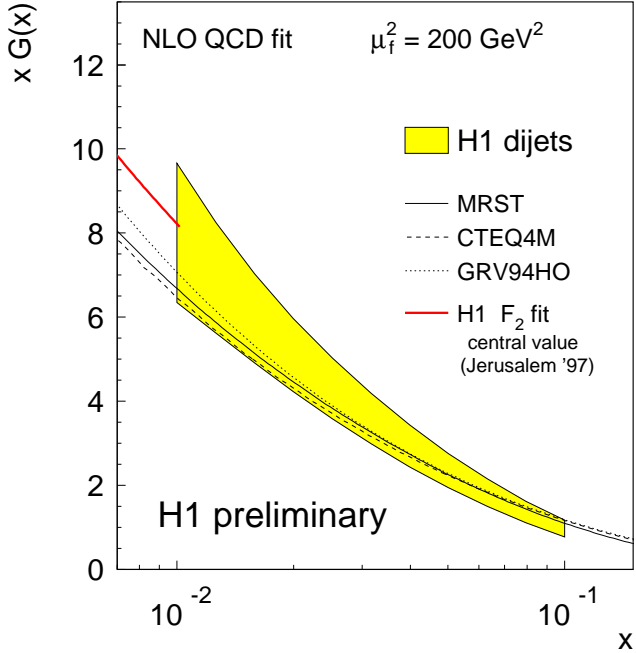,width=8.65cm}
\caption{The gluon density in the proton in the $\overline{MS}$-scheme
determined in a NLO QCD fit to 
dijet cross sections for a value of $\alpha_s(M_Z) = 0.119 \pm 0.005$. 
The error band is compared to the gluon densities obtained in global fits
and to the result of a QCD fit to H1 structure function data 
(only the central value without error band).}
\label{fig:gluon}
\end{figure}

The gluon density obtained from the QCD fit with its error band is shown 
in Fig.~\ref{fig:gluon} in the range $0.01 < x < 0.1$.
This result extends the $x$-range from previous gluon determinations from
H1 structure function data to larger $x$-values.
The uncertainty is dominated by the error of $\alpha_s$, the renormalization
scale dependence of the NLO calculation and the uncertainty in the 
absolute calorimetric energy scale.
While the result is slightly higher than the results of different 
global analyses (although compatible within the error)
it is in good agreement with the gluon density from a QCD analysis 
of H1 structure function data.

\section*{Summary}

Measurements of jet observables in hadronic final states in
deep-inelastic scattering and in hadron-hadron collisions have been presented.
Based on perturbative QCD calculations in next-to-leading order,
fits to the data have been performed to extract the value of the 
strong coupling constant $\alpha_s(M_Z)$ and the gluon density in the proton.
The results are consistent with each other and in agreement with
world average values.

The precision of these results is limited by uncertainties in the 
perturbative predictions (renormalization scale dependence) and by
the correlation between the gluon density and $\alpha_s$.
Further progress will require combined analyses of different data sets 
for consistent, simultaneous determinations of $\alpha_s$ and the 
gluon density.

The ``power correction" approach gives a fair description
of hadronization corrections to DIS event shape data.
Further developments in this field will improve the understanding
of non-perturbative physics and may allow to describe non-perturbative
contributions to different processes in a universal way.

\end{document}